\UseRawInputEncoding
\documentclass[%
 reprint,
 amsmath,amssymb,
 aps, showkeys
]{revtex4-2}
\usepackage{amsmath,amssymb,amsthm}
\usepackage{graphicx}
\usepackage{dcolumn}
\usepackage{bm}
\usepackage{hyperref}
\usepackage{comment}
\usepackage{tikz}

\usepackage{braket}
\usepackage{siunitx}
\usepackage[caption=false]{subfig}

\begin{document}

\preprint{APS/123-QED}

\title{Optimal photon budget allocation in E91 protocol}

\author{Melis Pahal{\i}}
\email{melis.pahali@ozu.edu.tr}
\affiliation{Electrical and Electronics Engineering Department, \"Ozye\u{g}in University, Istanbul 34794, Turkey}
\author{Kadir Durak}
\email{kadir.durak@ozyegin.edu.tr}
\affiliation{Electrical and Electronics Engineering Department, \"Ozye\u{g}in University, Istanbul 34794, Turkey}
\author{Utku Tefek}
\email{u.tefek@adsc-create.edu.sg}
\affiliation{Advanced Digital Sciences Center, Singapore 138602, Singapore}

\date{\today}

\begin{abstract}

In order for the deployment of quantum communication technologies in a global scale, it is necessary to meet data exchange demand of various size establishments per certain time intervals and it is important to make them cost effective and feasible. In order to meet these requirements, a lot of effort has been put into increasing key rate and having optimized systems. In this study, we focus on the improvement of raw key rate in a standardized E91 QKD system without compromising its security. Our method is to optimize photon budget allocation among three types of bits used in the system for different purposes or occurred unavoidably. These three types of bits are key bits, Bell's inequality bits and discarded bits and their ratios were $25\%$, $50\%$ and $25\%$, respectively, in the proof-of-principle experiment. On the other hand, we present $226.22\%$ increase in raw key rate with $83\%$, $10\%$ and $7\%$ allocation of photon budget among three types of bits. These ratios are achieved by replacing $50:50$ beam splitters with a $90:10$ beam splitter at one communicating side and a $93:7$ beam splitter at the other communicating side. Additionally, we demonstrate that the optimum beam splitting ratios can vary depending on photon budget. 
\end{abstract}

\maketitle


\section{\label{sec:intro}INTRODUCTION}

In quantum key distribution (QKD) literature, various methods were studied to increase key rate. For example, encryption and decryption in various degrees of freedoms \cite{Karimi2020IncreasingTR} were put into use, brightness of entanglement source was increased \cite{Ecker2020StrategiesFA} via fidelity and repetition rate increase, error correction with high speed \cite{HighSpeedAndAdaptableErrorCorr} and varied schemes \cite{ImprKeyRateOptFibQKD} were performed, high dimensional QKD protocols \cite{Ding2016HighdimensionalQK, TwoDDistributedPhaseReference} were generated, alternative QKD schemes like twin-field QKD \cite{OvercomingTheRateDistanceLimit, PhysRevX.8.031043, TwinFieldQKDoverA511km, Minder2019ExperimentalQK} were created and achievable distances were improved \cite{PhysRevLett.94.150501, LongdistanceBBM92overopticalfiber, MAFU2021e01008}.

In discrete-variable QKD, there are two entanglement-based protocols E91 \cite{QuantumcryptographybasedonBellstheorem} and BBM92 \cite{QuantumcryptographywithoutBellstheorem}, and there are different protocols which use entangled photons \cite{EntangledPhotonSixstatequantumcryptography, DIentanglementbasedBennett1992}. The studies performed related to E91 are security analysis \cite{ComparisonandAnalysisofBB84andE91QuantumCryptographyProtocolsSecurityStrengths, securityAnalysisOfE91protocolInCollectiveRotationNoise}, examination of protocol under a certain noise model \cite{securityAnalysisOfE91protocolInCollectiveRotationNoise}, employing optoelectronic polarization rotators (OPRs) in the implementation of violation of Bell's inequality \cite{EntanglementDemonstrationOnBoard} or throughout the QKD system \cite{ASystemANDMETHODFORQKDWITHENTANGLEDPHOTONPAIRS} as dynamic components instead of employing beam splitters and wave plates \cite{PhysRevA.78.020301} as the conventional static components, usage of entanglement between different degrees of freedoms (hybrid entanglement) \cite{ModifiedE91ProtocolDemonstrationWithHybridEntanglement}, efficiency analysis and improvement \cite{ImprovingefficiencyofQKDwithprobabilisticmeasurements}, feasibility study on satellite-based communication \cite{FeasibilityOfSatelliteQKD}, side channel analysis \cite{CryptographicSecurityConcernsonTimestampSharing}, and eavesdropping attack \cite{HackingtheBelltestusingclassicallightinenergytimeentanglementbasedQKD}.

Here, we studied E91 protocol. We focused on the key rate after the steps of sifting and parameter estimation, before the steps of error correction and privacy amplification, which can also be called raw key rate, and aimed at increasing it.

In the entanglement-based QKD modified protocol of Ekert which is studied in Ref. \cite{PhysRevA.78.020301}, there are three types of bits whose ratios are determined according to the beam splitting ratios of beam splitters at Alice's and Bob's sides. The three types of bits are key bits, Bell's inequality bits and discarded bits whose ratios are $25\%$, $50\%$ and $25\%$, respectively. These ratios are not optimum and our method of increasing the key rate is to optimize these ratios.

Photon budget defines the number of photons entering the detection module of a communicating side in a certain time interval. While the photon budget can take various values, a typical value is $10^6/\si{\second}$, as also assumed and used in our study.

For this study, our motivation is to increase the ratio of key bits in order to maximize key rate given photon budget without compromising the security of key. This can be achieved by decreasing the ratios of discarded bits and Bell's inequality bits. We examined various beam splitting ratios of beam splitters at Alice's and Bob's sides, and determined the optimum ratio of key bits yielding the maximum key rate. However, this optimization study can also be implemented to the QKD systems employing OPRs which are used for the same function as beam splitters and wave plates. This would be done by determining OPRs' optic axes' directions and the speed of change of the optic axes. Furthermore, this optimization study can also be implemented to the QKD systems which would use metamaterials having quantum phase transition materials for polarization change to be used in faster cryptography. The security of raw key relies on quantum bit error rate and Bell's inequality parameter. Quantum bit error rate will be a threshold determining the continuation or termination of the communication system. Bell's inequality parameter which depends on the amount of Bell's inequality bits will quantify how secure the system would be. 

Organization of the paper is the following. In Sec. \ref{sec:pvalue}, we investigated the uncertainty of the violation of Bell's inequality dependence to the number of photons used in the calculation of it and to the beam splitting ratios. In Sec. \ref{sec:DeadTime}, we examined the effect of dead time of a detector in photon counting and examined the concept on which photon budget allocation in the E91 protocol \cite{PhysRevA.78.020301} is based. In Sec. \ref{sec:AveKeyRate}, we introduced key rate expression that we use in our study and introduced the optimized photon budget. In Sec. \ref{sec:PDEandDC}, we presented the effects of photon detection efficiency and dark count of a detector on photon counting, explicitly. Lastly, in Sec. \ref{sec:Conc}, we summarize the topic by giving our results and mention the importance of the study.

\section{\label{sec:pvalue}Dependencies of UNCERTAINTY OF VIOLATION OF BELL'S INEQUALITY}

In the process of obtaining a final raw key to be used in the encryption and decryption, key bits and Bell's inequality bits are collected for a period of time. At the end of each period, the security of key bits is quantified with a parameter, which is generated by using Bell's inequality bits. If this parameter, so called Bell's inequality parameter, is below a threshold, that block of key bits cannot contribute to final raw key and is discarded. Bell's inequality parameter is calculated frequently within the process of key distribution until final raw key could be obtained. Whenever the parameter is above the threshold, the related block of key bits can be used to contribute final raw key. In this way, the blocks of key bits are collected and final raw key is obtained.

\begin{figure}[h!]
\subfloat(a)\hspace{45px}{
\begin{tikzpicture}[thick,scale=0.8, every node/.style={scale=0.9}]
\draw[style=help lines] (0,0) (3,2);

\coordinate (vec5) at (22.5:2); 
\coordinate (vec3) at (67.5:2);
\coordinate (vec4) at (337.5:2); 
\coordinate (vec6) at (292.5:2);
\coordinate (vecH) at (0:2);
\coordinate (vecV) at (90:2);
\coordinate (vec_X) at (270:2);
\coordinate (vec_Y) at (180:2);

\draw[->,thick,brown] (0,0) -- (vec5) node[below right] {$5 (\ang{22.5})$};
\draw[->,thick,red] (0,0) -- (vec3) node[below right] {$3 (\ang{67.5})$};
\draw[->,thick,red] (0,0) -- (vec4) node[below right] {$4 (\ang{-22.5})$};
\draw[->,thick,brown] (0,0) -- (vec6) node[below right] {$6 (\ang{-67.5})$};
\node[above,font=\large\bfseries] at (current bounding box.north) {Alice}; 
\end{tikzpicture}   
\begin{tikzpicture}[thick,scale=0.8, every node/.style={scale=0.9}]
\draw[style=help lines] (0,0) (3,2);

\coordinate (vec3) at (45:2); 
\coordinate (vec4) at (-45:2);
\coordinate (vecH) at (0:2);
\coordinate (vecV) at (90:2);
\coordinate (vec_X) at (270:2);
\coordinate (vec_Y) at (180:2);

\draw[->,thick,black] (0,0) -- (vec3) node[below right] {$3' (\ang{45})$};
\draw[->,thick,black] (0,0) -- (vec4) node[below right] {$4' (\ang{-45})$};
\draw[->,thick,blue] (0,0) -- (vecH) node [below] {$2'(H)$};
\draw[->,thick,blue] (0,0) -- (vecV) node [left] {$1'(V)$};

\node[above,font=\large\bfseries] at (current bounding box.north) {Bob};     
\end{tikzpicture}}
\subfloat(b){\includegraphics[width=8cm]{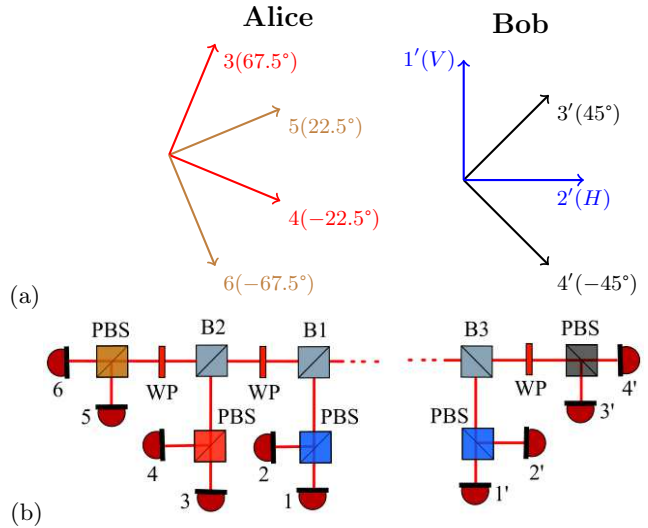}}
\caption{(a) Measurement basis sets used in the calculation of the amount of violation of Bell's inequality at Alice's and Bob's sides. The four basis sets and their components are as follows: $a_0\equiv\{3,4\}$, $a_1\equiv\{5,6\}$, $b_0\equiv\{1',2'\}$, $b_1\equiv\{3',4'\}$. (b) B1, B2, B3: Beam splitters. PBS: Polarizing beam splitter. WP: Wave plate. The numbers indicate detectors, in other words, the basis sets' components. Figure on the left and right represent the detection modules at Alice's and Bob's sides, respectively. Reflected outputs of B1 and B3 constitute key bit detection parts of the detection modules.}
\label{fig:BellBazlari}
\end{figure}

The Bell's inequality parameter we used throughout the chapter is given in Eq.~(\ref{eq:S}).
\begin{equation}\label{eq:S}
    S=E(a_0,b_0)+E(a_0,b_1)-E(a_1,b_0)+E(a_1,b_1)
\end{equation}
where $a_0$($b_0$) and $a_1$($b_1$) are two sets of detectors located at Alice's (Bob's) detection module. Each set of detectors consists of two detectors. In other words, each of $a_0$($b_0$) and $a_1$($b_1$) can be called a basis set and consists of two orthogonal basis. Thus each detector is related to a basis and is dedicated to measure a polarization direction which is shown in Fig. \ref{fig:BellBazlari} \cite{PhysRevA.78.020301}. $E(a_i,b_i)$ is a correlation coefficient related to measurement results obtained from basis sets $a_i$ and $b_i$. Lastly, $S$ is the combined correlation coefficient used to quantify the quantumness a QKD system has. While the Bell's inequality satisfies $|S|\leq2$, which we call classical region, its violation is realized when $2<|S|\leq2\sqrt{2}$. 

In a pair of measurement among Alice and Bob, the coincidence of two of the basis in Fig. \ref{fig:BellBazlari} (a) results in a Bell's inequality bit being obtained in the QKD system. The reflected outputs of B1 and B3 in Fig. \ref{fig:BellBazlari} (b) represent the same basis set, thus the coincidence of these basis sets results in a key bit being obtained in the system. Lastly, the coincidence of a measurement at the reflected output of B1 and a measurement at the basis set labeled with $\{3',4'\}$ results in a discarded bit being obtained in the system. Namely, the type of a bit is determined by which basis sets coincide.

The total number of photons which is used in the calculation of $S$ is allocated to the components of $S$, $n_{kl}$. The mathematical expression of it is given in Eq.~(\ref{eq:nklSum}).
\begin{equation}\label{eq:nklSum}
    \eta=\sum_{k}\sum_l n_{kl}
\end{equation}
where $\eta$ is the total number of photons used in the calculation of $S$, $n_{kl}$ is the components of $S$ with $k\equiv\{3,4,5,6\}$ and $l\equiv\{1',2',3',4'\}$ shown in Fig. \ref{fig:BellBazlari} (b). $n_{kl}$ can be written in terms of two independent parameters, which are $\eta$ and the reflectance of the beam splitter B3 in Fig. \ref{fig:BellBazlari} (b). Thus, it can be interpreted as $S$ is a function of independent and random variables. Propagation of uncertainty can be applied to $S$ and the uncertainty of $S$ can be expressed as given in Eq.~(\ref{eq:PropOfUn}) \cite{ApproachingtheTsirelsonboundwithaSagnacsourceofpolarizationentangledphotons, TowardsaLoopholeFreeTestofBellsInequalitywithEntangledPairsofNeutralAtoms}. 
\begin{equation}\label{eq:PropOfUn}
    \delta S=\sqrt{\sum_{k}\sum_{l}\Big(\frac{\partial S}{\partial n_{k,l}}\delta n_{k,l}\Big)^2}
\end{equation}
where $\delta$ stands for uncertainty. The definition of uncertainty of coincidence counts is given in Eq. (\ref{eq:uncertainty}). Uncertainty of $S$, $\delta S$, can give an idea about how reliable an $S$ would be relatively. Lastly, $\partial$ stands for partial derivative.
\begin{equation}\label{eq:uncertainty}
    \delta n_{k,l}=\sqrt{n_{k,l}}
\end{equation}

\begin{figure}[h!]
    \centering
    \includegraphics[width=263px]{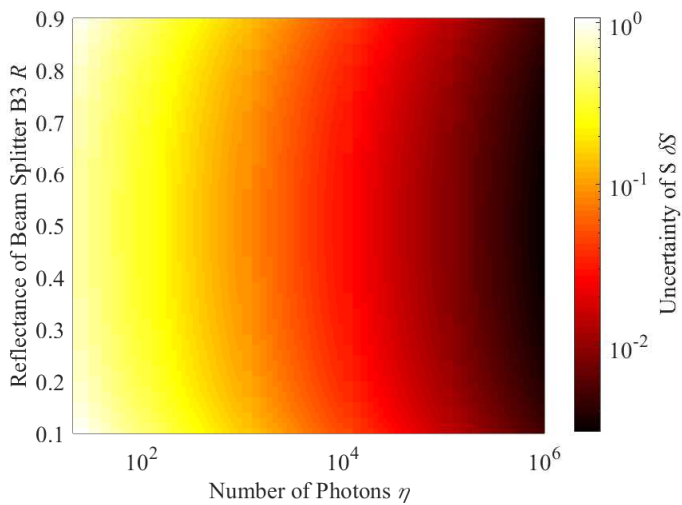}
    \caption{The graph of the uncertainty of $S$ dependence to the number of photons and the reflectance of beam splitter B3. Color bar represents the uncertainty of $S$.}
    \label{fig:DependenceOfStoNSandR} 
\end{figure}

A mixed state consisting of singlet state $\ket{\psi^-}=\frac{1}{\sqrt{2}}(\ket{01}-\ket{10})$ having visibility $0.95$ and white noise, which is also called Werner state, was used in the quantum channel. For the mixed state $\rho$, the uncertainty of $S$ depending on the various numbers of photons and various reflectance of beam splitter B3 was computed and shown in Fig. \ref{fig:DependenceOfStoNSandR}.

\begin{figure}[h!]
    \centering
    \includegraphics[width=263px]{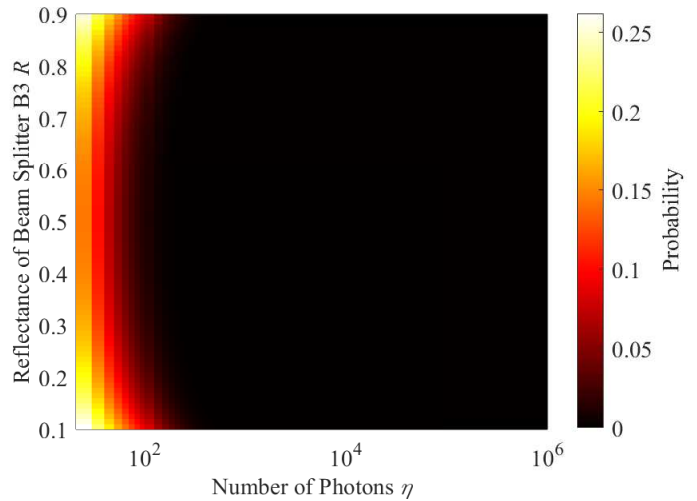}
    \caption{The graph of the dependence of the probability to the number of photons and the reflectance of beam splitter B3. Color bar represents the probability of obtaining a value from the classical region for $S$.}
    \label{fig:ProbVsRandEta}
\end{figure}

We considered different numbers of photons as independent experiments and modeled the probability distribution of $S$ in each case with a Gaussian distribution approximation whose standard deviation is the uncertainty of $S$ in Fig. \ref{fig:DependenceOfStoNSandR}. We called the region $|S|\leq2$ as classical region, computed the areas under the distributions which fall into the classical region and showed the dependence of the probability to the number of photons and reflectance of beams splitter B3 in Fig.\ref{fig:ProbVsRandEta}.

The probability shows the percentage of all the blocks of key bits collected in a key distribution process which cannot contribute to the final raw key. In other words, it represents the ratio of key rate which will be discarded in unit time interval. Even if the quantum channel is secure, communicating sides can obtain an $S$ value from the classical region at that probability.  We will call the key rate remains after the decrease at that probability as average key rate in the rest of the paper.

\section{\label{sec:DeadTime}DEAD TIME EFFECT IN PHOTON COUNTING}

\begin{figure}[h!]
    \centering
    \includegraphics[width=267px]{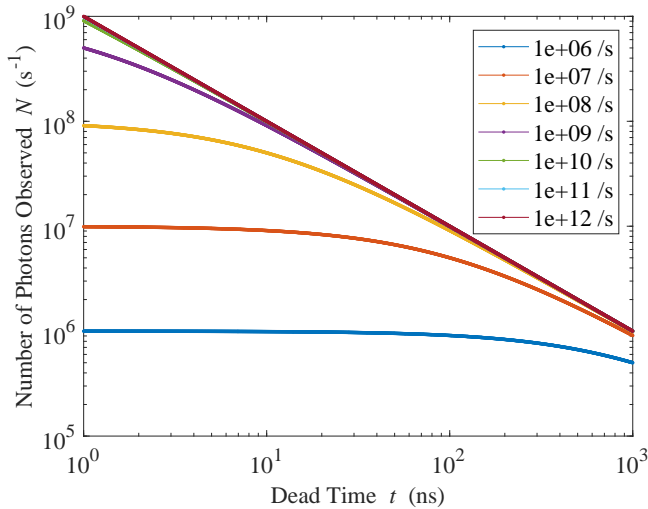}
    \caption{The graph of the number of photons observed at a detector per second dependence to the dead time of detector. The legend represents $n$, namely, the number of photons reaching detector per second. The curves from bottom to top correspond to $n$ from $1e+06/\si{\second}$ to $1e+12/\si{\second}$.}
    \label{fig:BW}
\end{figure}

Dead time is the bottleneck which limits photon counting in a detector. The number of photons observed at a detector may differ from the number of photons reaching the detector. The relation between them is given in Eq.~(\ref{eq:BW}).
\begin{equation}\label{eq:BW}
    N=\frac{n}{n\cdot t\cdot 10^{-12}+1}
\end{equation}
where $n$ is the number of photons reaching detector per second, $t$ is the dead time of detector in terms of $\si{\pico\second}$ and takes value within the range of $1\si{\pico\second}\leq t\leq 10^{12}\si{\pico\second}$, and $N$ is the number of photons observed at detector per second. Dependence of $N$ to $t$ is shown in Fig. \ref{fig:BW}. Dead time changes within the range of $t=[1\si{\nano\second},10\si{\micro\second}]$ because of presenting the most common values encountered in active and passive quenching detectors. 

There are three types of $n$ with which we are dealing. Photon budget is allocated among three types of $n$. They are the numbers of photons reaching detectors or detectors' groups which are dedicated to obtain three types of bits. The numbers of photons reaching these detectors are determined by the ratios of three types of bits. $N_{key}$ will the number of photons observed per second and will be used in the calculation of key rate. In the rest of the paper, dead time was be taken as $1\si{\pico\second}$. Given the dead time, we continued our study with the numbers of photons observed in three types of bits.

\section{\label{sec:AveKeyRate}AVERAGE KEY RATE}

The key rate formula we used is that of device-independent QKD which is based on collective attacks scenario \cite{DIQKDsec}. It is the first multiplier in Eq. (\ref{eq:KeyRate}) whose unit is bits/symbol.

\begin{equation}\label{eq:KeyRate}
    \Bigg(1-h(Q)-h\Big(\frac{1+\sqrt{((S+\delta S)/2)^2-1}}{2}\Big)\Bigg)\cdot N_{key}\cdot (1-P)
\end{equation}
where $h(Q)$ is the binary entropy of quantum bit error rate $Q$, $h\Big(\frac{1+\sqrt{((S+\delta S)/2)^2-1}}{2}\Big)$ is the accessible information about the quantum channel by an eavesdropper where $S$ is the mean and $\delta S$ is the uncertainty of Bell's inequality parameter. In the comparison of different distributions of $S$ in the experiments using different total number of photons, the range of $\mu\pm\sigma$, where $\mu$ is the mean and $\sigma$ is the standard deviation of the distribution, can be used. Thus from the range of $S\pm\delta S$, we chose $S+\delta S$ because of being closer to the classical region to distinguish the effect of different distributions of $S$ as the worst case scenario that can be encountered in an experiment. $N_{key}$ is as defined in Sec. \ref{sec:DeadTime} and is the conversion parameter from bits/symbol to bps which would be the unit of the key rate used in the rest of the paper since each symbol or photon carries one bit of information. $P$ is the probability of obtaining an $S$ from the classical region and with the use of it the key rate formula turns into average key rate.

\begin{figure}[h!]
    \centering
    \includegraphics[width=267px]{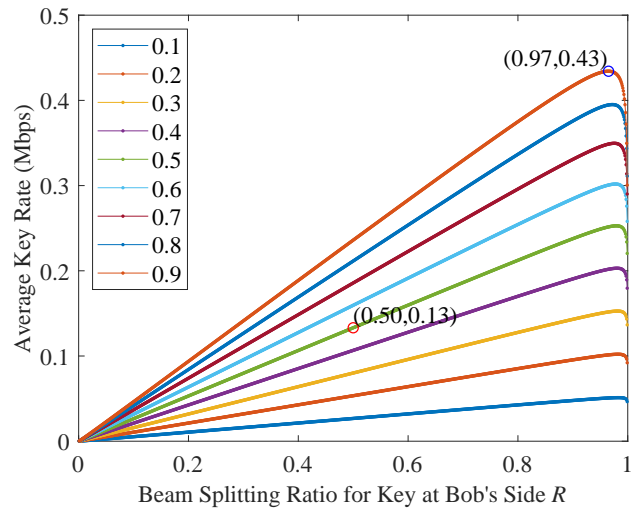}
    \caption{The graph of the average key rate dependence to the beam splitting ratio for key at Bob's side, namely, dependence to the beam splitting ratio at the reflected output of B3 in Fig. \ref{fig:BellBazlari} (b). The legend represents the beam splitting ratio for key at Alice's side, namely, the beam splitting ratio at the reflected output of B1 in Fig. \ref{fig:BellBazlari} (b). The curves from bottom to top correspond to the beam splitting ratio for key at Alice's side from $0.1$ to $0.9$.}
    \label{fig:AveKey2}
\end{figure}

\begin{figure}[h!]
    \centering
    \includegraphics[width=250px]{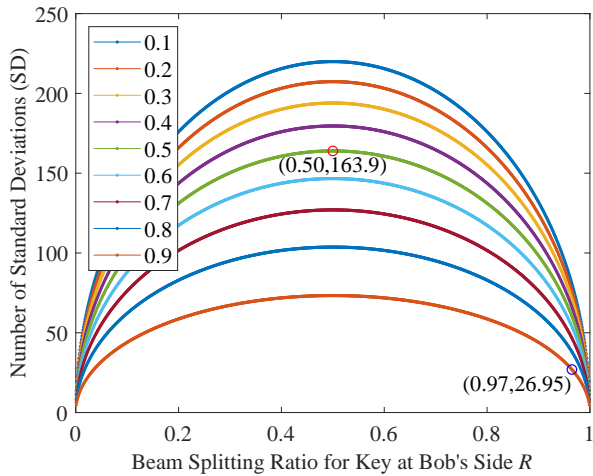}
    \caption{The graph of the number of standard deviations of the violation of Bell's inequality dependence to the beam splitting ratios for key. The legend represents the beam splitting ratio for key at Alice's side.}
    \label{fig:NumSD}
\end{figure}

The average key rates that can be obtained depending on the beam splitting ratios, and the maximized key rate can be seen in Fig. \ref{fig:AveKey2}. While average key rate is $0.1331$Mbps in the standard configuration which is the employment of $50:50$ beam splitters \cite{PhysRevA.78.020301} and is shown as a red circle in the figure, it becomes $0.4342$Mbps in the employment of $90:10$ beam splitter at Alice's side and $97:3$ beam splitter at Bob's side and shown as a blue circle in the figure. Hence our method presents an increase in the average key rate up to $226.22\%$. Also, while the ratios of key bits, Bell's inequality bits and discarded bits are $25\%$, $50\%$ and $25\%$ in the standard configuration, their ratios in the maximized key rate configuration is $87\%$, $10\%$ and $3\%$. Additionally, the number of photons used in the calculation of $S$ becomes $10^5/\si{\second}$ with the $10\%$ allocation. $S$ with its uncertainty corresponding to that much of photons is $-2.6870\pm0.0255$. Bell's inequality is violated with $26.95$ standard deviations (SD), it can be seen in Fig. \ref{fig:NumSD}. In the literature, there are various numbers of standard deviations in the violations of Bell's inequality, such as $6$SD, $12$SD \cite{BellinequalitiesforentangledqubitsQuantitativetestsofquantumcharacter}, $8.5$SD, $9.3$SD \cite{PMID:28490728}, $10$SD \cite{PhysRevLett.64.2495, Tang_2022}, $24$SD \cite{Ma2009ExperimentalVO, Tang_2022}, $203$SD \cite{Extremeviolationoflocalrealismwithahyper-entangledfour-photon-eight-qubit}, and there is no threshold other than the trivial conclusion that the number of standard deviations $>0$ must be satisfied. Thus $26.95$SD is within the acceptable range of the numbers of standard deviations of the violation, and $10\%$ allocation instead of $50\%$ to the Bell's inequality bits does not expose a compromise in the security of the QKD system which is based on $S$.

Incremental steps of the beam splitting ratio at Alice's side is $0.1$ and at Bob's side is $0.001$. More precise optimum values can be obtained by reducing the incremental steps of the beam splitting ratios in the simulations for the QKD systems which enable more sensitive implementations. However, the change in the beam splitting ratios from $50:50$ to roughly $90:10$ would give a considerable increase in the key rate as shown in this study.

\begin{figure}[h!]
    \centering
    \includegraphics[width=267px]{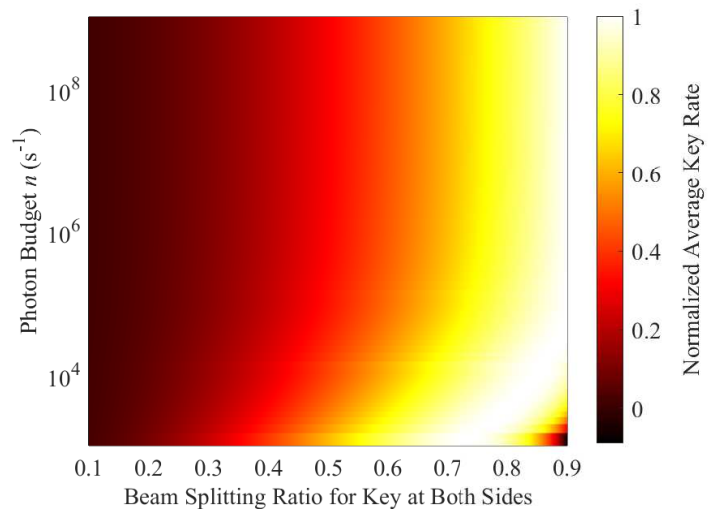}
    \caption{Normalized average key rate dependence to photon budget and beam splitting ratios for key at both sides.}
    \label{fig:NormalizedAve}
\end{figure}

Fig. \ref{fig:NormalizedAve} is plotted to show at which beam splitting ratios a maximized average key rate can be achieved. In this figure, the beam splitting ratios at both sides are chosen as the same as indicated in the label of x-axis. Photon budget changes within the range of $[10^3, 10^9]/\si{\second}$. Color bar shows normalized average key rate in which average key rates at various beam splitting ratios for a photon budget is normalized according to the maximized average key rate. After $n=10^6/\si{\second}$, maximized average key rate seems to happen when beam splitting ratios for key $=0.9$. It is suggested to look at average key rate change in a graph of the type of Fig. \ref{fig:AveKey2}. If the global maximum average key rate cannot be seen in such a graph, the conclusion that this experimental setup for higher photon budgets may not be effective can be inferred.

\section{\label{sec:PDEandDC}Photon Detection Efficiency and Dark Count Effects in Photon Counting}

\begin{figure}[h!]
\subfloat(a){\includegraphics[width=255px]{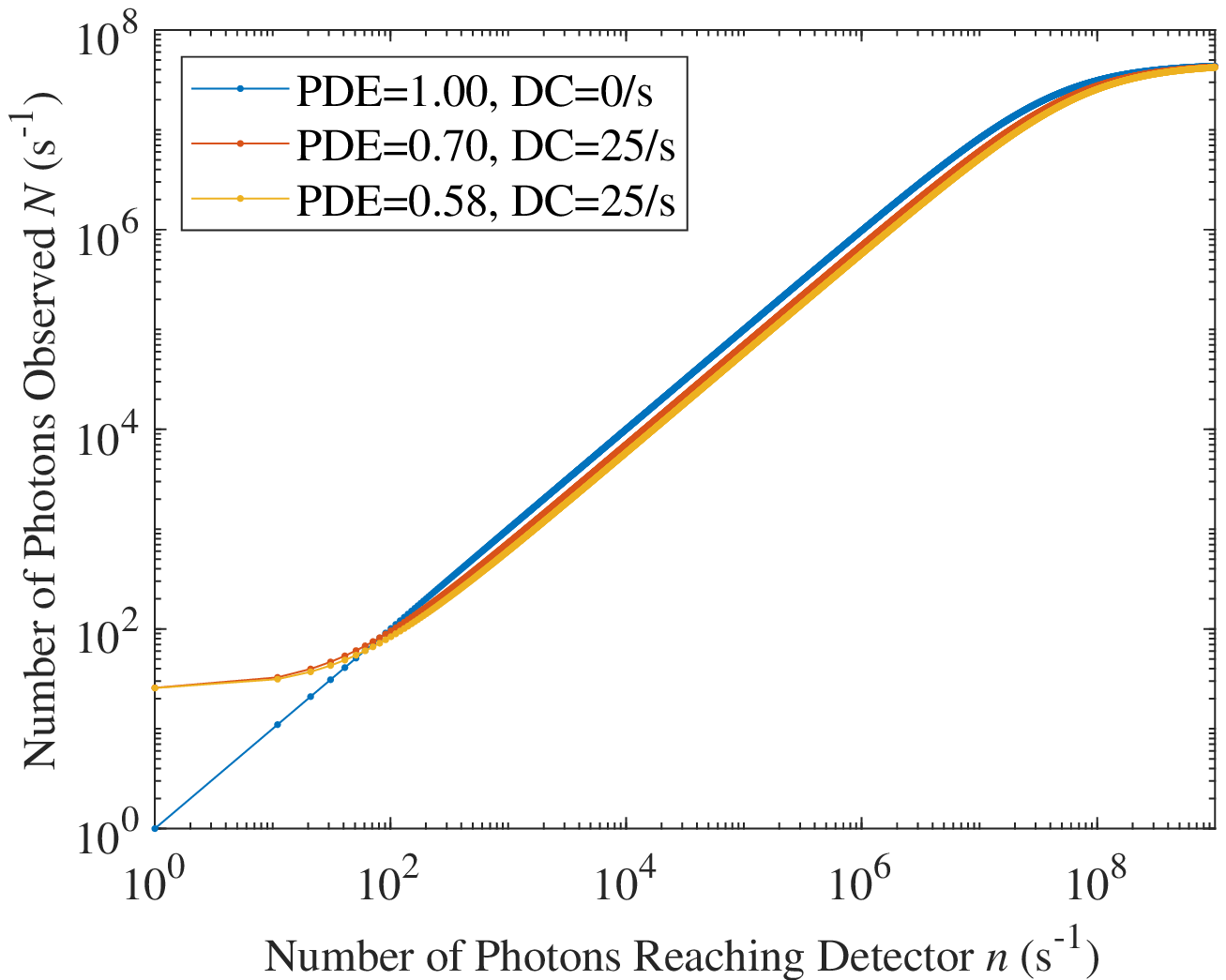}}
\subfloat(b){\includegraphics[width=255px]{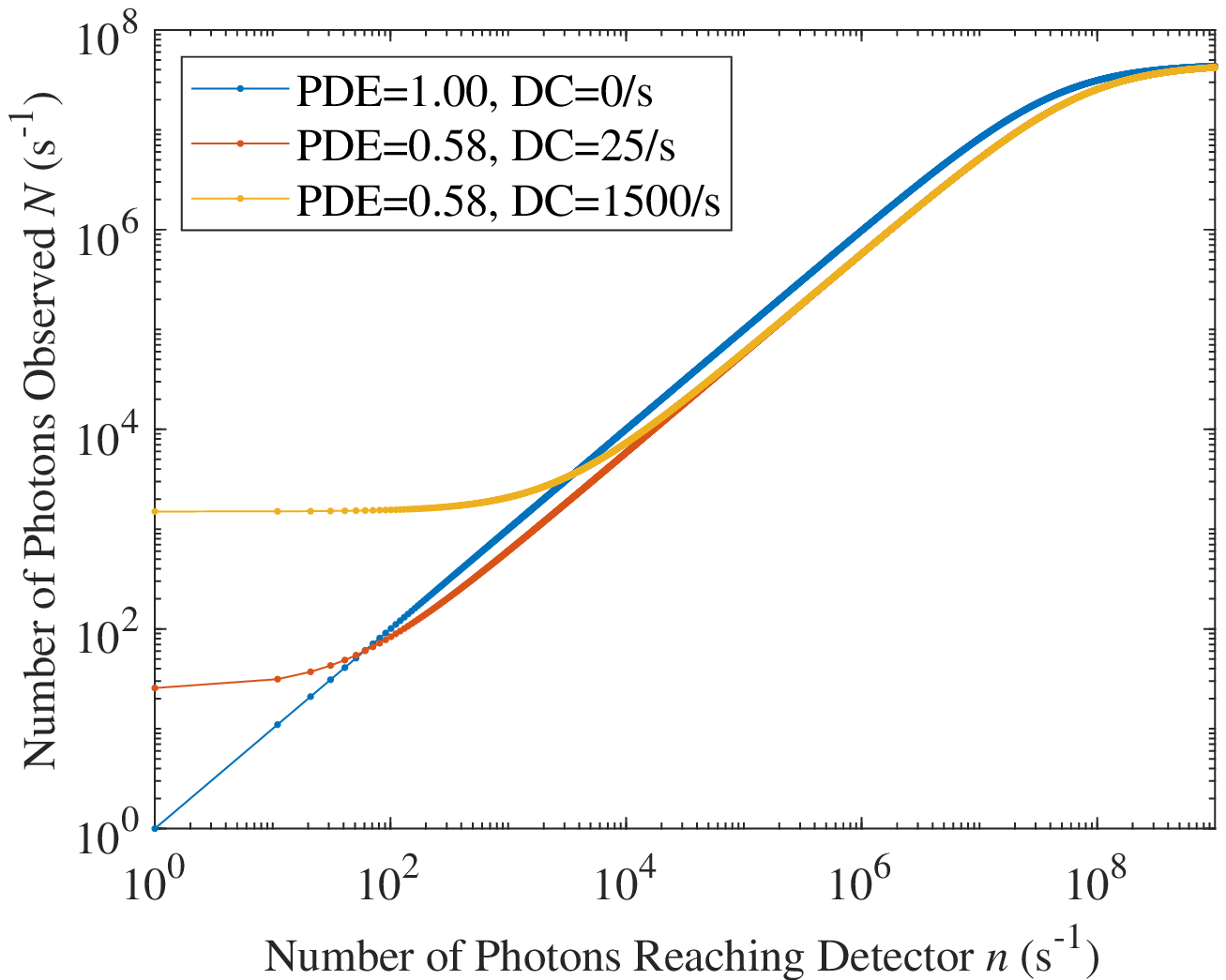}}
\caption{The graphs of the number of photons observed vs the number of photons reaching detector. The effect of photon detection efficiency and dark count to the number of photons observed.}
\label{fig:PDandDCEeffect}
\end{figure}

Besides dead time, photon detection efficiency (PDE) and dark count (DC) also take part in the relation between the numbers of photons reaching a detector and observed at the detector per second. They can be incorporated as given in Eq.(\ref{eq:PDE}).
\begin{equation}\label{eq:PDE}
    N=\frac{n\cdot PDE+DC}{n\cdot PDE\cdot t\cdot 10^{-12}+1}
\end{equation}
where $n$, $t$ and $N$ are as defined in Eq. \ref{eq:BW}. In this study, PDE and CD are taken as $1$ and $0/\si{\second}$, respectively, for the simplification. The effect of PDE can be seen in the left hand side of Fig. \ref{fig:PDandDCEeffect}, which is a vertical shift in the linear region. And the effect of DC can be seen in the right hand side of Fig. \ref{fig:PDandDCEeffect}, which shows itself as noise while working with relatively lower number of photons. And it shows the performance of a detector while working with relatively low number of photons. In this figure, dead time was taken as $22\si{\nano\second}$ for all the curves and substituted into Eq.\ref{eq:PDE} in terms of $\si{\pico\second}$ as compatible with Eq.\ref{eq:BW}. Alternative to Eq.(\ref{eq:PDE}), a formula which characterizes a detector should be used in a specific implementation in order to obtain average key rates achievable.

\section{\label{sec:Conc}CONCLUSION}

In summary, we focused on a standardized E91 QKD protocol, in which $50:50$ beam splitters are employed at both Alice's and Bob's sides. Depending on the positioning of measurement basis sets with respect to the beam splitters, there was the allocation of $25\%$, $50\%$ and $25\%$ of a photon budget among three types of bits, which are key bits, Bell's inequality bits and discarded bits. We calculated raw key rate for the $50:50$ beam splitters and for the standard allocation as a reference point. Then, we examined the value of the key rate for various beam splitting ratios and for various allocations of the photon budget. We searched the optimum beam splitting ratios and allocations yielding the global maximum key rate, and showed that $226.22\%$ increase in key rate can be achievable with $90:10$ and $97:3$ beam splitting ratios, and with the allocation of $87\%$, $10\%$ and $3\%$. Additionally, we showed that optimum beam splitting ratios can vary for different photon budgets in a constant configuration of experimental setup by defining a parameter called normalized average key rate and by using it. An E91 QKD protocol may have various configurations such as not including all the basis sets used in this study or establishing symmetrical detection modules in terms of basis sets at both communicating sides or the realization of basis sets by using OPRs instead of beam splitters and wave plates. The optimization of photon budget allocation should be applied to all the experimental setups of the similar kind for the best use of resources and for the maximization of key rate.

\section*{ACKNOWLEDGMENT}

This study was supported by the Scientific and Technological Research Council of Turkey (T\"{U}B\.{I}TAK) Academic Research Funding Programs Directorate (ARDEB), Primary Subjects R\&D Funding Program (Program No. 1003), Project No. 118E991.

\newpage
\bibliography{apssamp}

\end{document}